\documentstyle[epsfig,mathptm]{mn} 
\newif\ifAMStwofonts
\AMStwofontstrue                
\DeclareMathVersion{bold}                    
\DeclareMathAlphabet{\mathbfit}{OT1}{cmr}{bx}{it}
\SetMathAlphabet\mathbfit{bold}{OT1}{cmr}{bx}{it}
\DeclareMathAlphabet{\mathbfss}{OT1}{cmss}{bx}{n}
\SetMathAlphabet\mathbfss{bold}{OT1}{cmss}{bx}{n}
\ifAMStwofonts
  \ifCUPmtlplainloaded \else
    \DeclareSymbolFont{UPM}{U}{eur}{m}{n}
    \SetSymbolFont{UPM}{bold}{U}{eur}{b}{n}
    \DeclareSymbolFont{AMSa}{U}{msa}{m}{n}
    \DeclareMathSymbol{\upi}{0}{UPM}{"19}
    \DeclareMathSymbol{\umu}{0}{UPM}{"16}
    \DeclareMathSymbol{\upartial}{0}{UPM}{"40}
    \DeclareMathSymbol{\leqslant}{3}{AMSa}{"36}
    \DeclareMathSymbol{\geqslant}{3}{AMSa}{"3E}
  \fi
\fi

\newcommand{\dg}{^\circ}
\title[Foreground separation for MAP observations]
{Separation of foregrounds from
cosmic microwave background observations with the MAP satellite} 
\author[A.W. Jones, M.P Hobson and A.N.~Lasenby]
{A.W.~Jones, M.P.~Hobson and A.N.~Lasenby \\
Cavendish Astrophysics, Cavendish Laboratory, 
Madingley Road, Cambridge CB3 OHE, UK}
\date{Accepted ???. Received ???; in original form \today}
\pagerange{\pageref{firstpage}--\pageref{lastpage}}
\pubyear{1998}

\begin{document}
\maketitle
\label{firstpage}

\begin{abstract}
Simulated observations of a $10\dg \times 10\dg$ field by the
Microwave Anisotropy Probe (MAP) are analysed in order to
separate cosmic microwave background (CMB) emission from
foreground contaminants and instrumental noise and thereby determine
how accurately the CMB emission can be recovered. The simulations
include emission from the CMB, the kinetic and thermal
Sunyaev-Zel'dovich (SZ) effects from galaxy clusters, as well as
Galactic dust, free-free and synchrotron.  We find that, even in the
presence of these contaminating foregrounds, the CMB map is
reconstructed with an rms accuracy of about 20 $\mu$K per 12.6 arcmin
pixel, which represents a substantial improvement as compared to the
individual temperature sensitivities of the raw data channels.  We
also find, for the single $10\dg \times 10\dg$ field, that the CMB
power spectrum is accurately recovered for $\ell \la 600$.
\end{abstract}


\section{Introduction} 
\label{intro}

The NASA MAP satellite is due to be launched in 2000 and aims to make
high-resolution, low-noise maps of the whole sky at several observing
frequencies. The main goal of the mission is to use this
multi-frequency data to produce an all-sky map of fluctuations in the
cosmic microwave background (CMB). As with any CMB experiment,
however, MAP is also sensitive to emission from several foreground
components. The main contaminants are expected to be Galactic dust,
free-free and synchrotron emission, the kinetic and thermal
Sunyaev-Zel'dovich (SZ) effect from galaxy clusters, and
extra-galactic point sources.  In order to obtain a map of the CMB
fluctuations alone, it is therefore necessary to separate the emission
from these various components.

The contamination due to extra-galactic point sources is expected to
be mainly from radio-loud AGNs, including flat spectrum radio-galaxies
and QSOs, blazars and possibly some inverted spectrum
radiosources. Since the frequency spectra of many of these
extra-galactic objects are, in general, rather complicated, any
extrapolation of their emission over a wide frequency range must be
performed with caution. For MAP observations it is possible that a
significant fraction of point sources may be identified and removed
using the satellite observations themselves, together perhaps with
pre-existing surveys.  Moreover, by including the point source
predictions of Toffolatti et al. (1998) into simulated Planck Surveyor
observations, Hobson et al. (1998b) find, using a maximum-entropy
algorithm, that the quality of the component separation is
relatively insensitive to the presence of the point sources.
Therefore, the effects of point sources will be ignored in this paper.

Aside from extra-galactic point sources, the other physical components
mentioned above have reasonably well defined spectral characteristics,
and we may use this information, together with multi-frequency
observations, to distinguish between the various foregrounds.  In this
paper, we perform a separation of the different components, in order
to determine the accuracy to which the MAP satellite can recover the
CMB emission in the presence of contaminating foreground emission.
The separation is carried out using the Fourier space maximum-entropy
method (MEM) developed by Hobson et al. (1998) (hereafter HJLB98),
which in the absence of non-Gaussian signals reduces to linear Wiener
filtering (Bouchet, Gispert \& Puget 1996; Tegmark \& Efstathiou
1996).

\section{Simulated MAP observations} 
\label{simmobs} 

\begin{figure*}
\centerline{\epsfig{
file=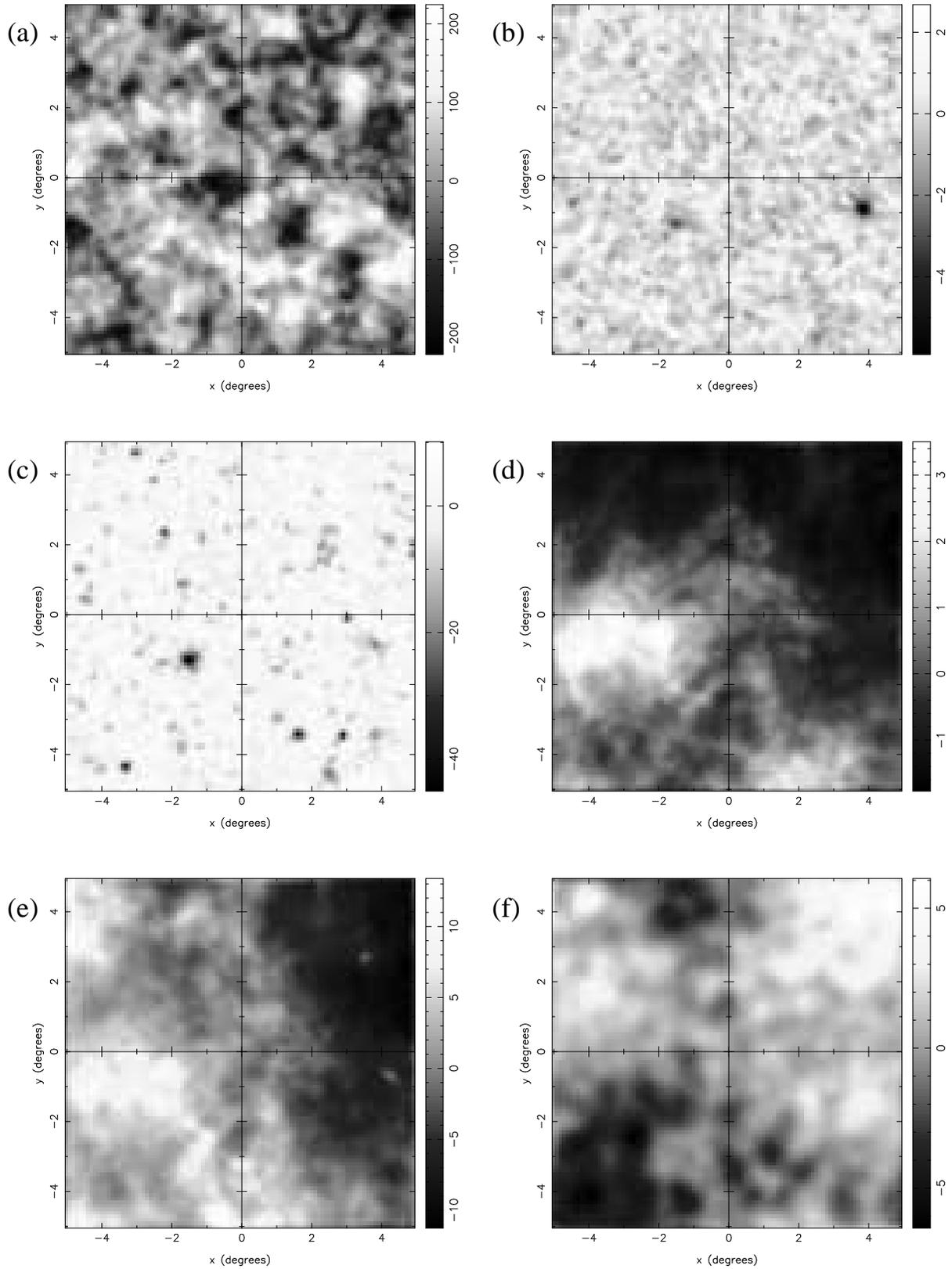,width=16cm}}
\caption{The $10\dg \times 10\dg$ realisations of the six input
components used to make simulated MAP observations:
(a) primary CMBR fluctuations; (b) kinetic
SZ effect; (c) thermal SZ effect; (d) Galactic dust; (e) Galactic
free-free; (f) Galactic synchrotron emission. 
Each component is plotted at 50 GHz
and has been convolved with a Gaussian beam of FWHM equal to 12.6
arcmin, the maximum angular resolution proposed for the MAP
satellite. The map units are equivalent thermodynamic temperature in
$\mu$K.}
\label{fig1}
\end{figure*}

The simulated input components used here are identical to those
described in HJLB98, so that a direct comparison between
the MAP and Planck Surveyor results can be made. These simulations
were performed by Gispert \& Bouchet (1997) 
and consist of a $10\dg \times 10\dg$
field for each of the six foreground components described above.
The realisations of the input components used are shown in
Figure~\ref{fig1}. Each component is plotted at 50~GHz and, for
illustration purposes, has been convolved with a Gaussian beam of
FWHM equal to 12.6 arcmin, which is the highest resolution of the
current design for the MAP satellite. For convenience, the mean of
each map is set to zero, in order to highlight the relative level of
fluctuations due to each component. 

The primary CMB fluctuations are a realisation of COBE-normalised
standard CDM with critical density and $H_\circ=50$km s$^{-1}$
Mpc$^{-1}$. IRAS 100-$\mu$m maps are used as spatial templates for the
dust and free-free components; the correlated emission from
these two components is described by HJLB98. 
Haslam 408~MHz maps (Haslam et al. 1982) are
used as the spatial template for the synchrotron emission to which was
added Gaussian small scale structure following a $C_\ell \propto
\ell^{-3}$ power law on angular scales below 0.85 degrees.
The kinetic and thermal SZ effects are generated using a Press-Schechter
formalism, as discussed in Bouchet et al. (1997) and a King model for
the cluster profiles. The MAP satellite is not designed to be
sensitive to either of the SZ effects but they are included here for
completeness. 

Using the realisations for each physical component and the 
design specifications summarised in Table~\ref{table1}, it is
straightforward to simulate MAP satellite observations. 
The simulated observations are produced by integrating the emission
due to each physical component across each waveband, assuming the
transmission is uniform across the band. At each observing frequency,
the total sky emission is convolved with a beam of the appropriate
FWHM. Finally, isotropic noise is added to the maps, assuming a
spatial sampling rate of FWHM/2.4 at each frequency. We have assumed
that any striping due to the scanning strategy and $1/f$ noise has
been removed to sufficient accuracy that any residuals become
negligible. 

\begin{figure}
\centerline{\epsfig{
file=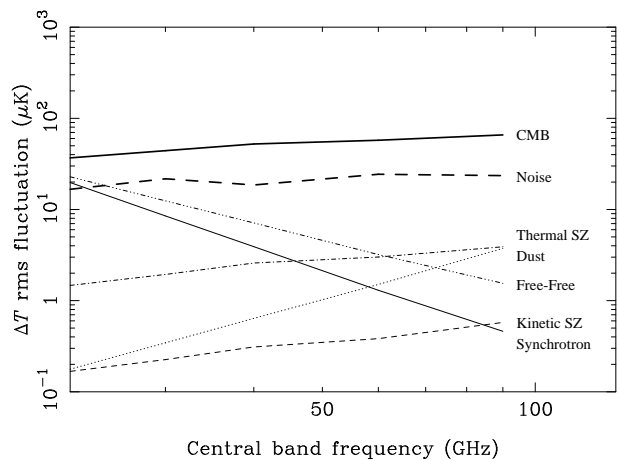,width=6cm,angle=270}}
\caption{The rms thermodynamic temperature fluctuations at the
MAP satellite 
observing frequencies due to each physical component, after convolution
with the appropriate beam and using a sampling rate of FWHM/2.4. The
rms noise per pixel at each frequency channel is also plotted.}
\label{fig2}
\end{figure}

Figure~\ref{fig2} shows the rms temperature fluctuations as a function
of observing frequency due to each physical component, after
convolution with the beam. The rms noise per pixel at each frequency
is also plotted (this noise is calculated for one year of observations
and is based upon the current design of the satellite as reported in
the MAP NASA home page). We see from the figure that, as expected, the rms
temperature fluctuation of the CMB is almost constant across the
frequency channels, the only variations being due to the convolution
with beams of different sizes. Furthermore, it is seen that the CMB is
consistently above the noise level, but only at the very lowest
frequency do any of the other physical components become
significant. At 22~GHz it is seen that the free-free and synchrotron
components just reach above the noise level but are still well below
the level of the CMB. This is to be expected since the MAP satellite is
designed not to make separate maps of the individual components, but 
only to provide sufficient information about
the foreground emission in order to perform an accurate
subtraction from the CMB. This should be contrasted with the
Planck Surveyor mission (Bersanelli et al. 1996), which is designed to produce
all-sky maps of the foreground components as well as the CMB.
\begin{table}
\begin{center}
\caption{Proposed observational parameters for the MAP
satellite. 
Angular resolution is quoted as FWHM 
for a Gaussian beam. Sensitivities are quoted 
for 12 months of observations.}
\begin{tabular}{lcccccc} \hline
Central frequency (GHz):   
& 22   & 30   & 40   & 60  & 90  \\
Fractional bandwidth ($\Delta\nu/\nu$):   
& 0.2 & 0.2 & 0.2 & 0.2 & 0.2 \\
Transmission:
& 1.0  & 1.0  & 1.0  & 1.0  & 1.0 \\
Angular resolution (arcmin):  
& 56   & 41   & 28   & 21   & 12.6 \\
$\Delta T$ sensitivity ($\mu$K) ($17^\prime$ pixel): 
& 26.  & 32.  & 27. & 35. & 35. \\
\hline
\end{tabular}
\end{center}
\label{table1}
\end{table}

\begin{figure*}
\centerline{\epsfig{
file=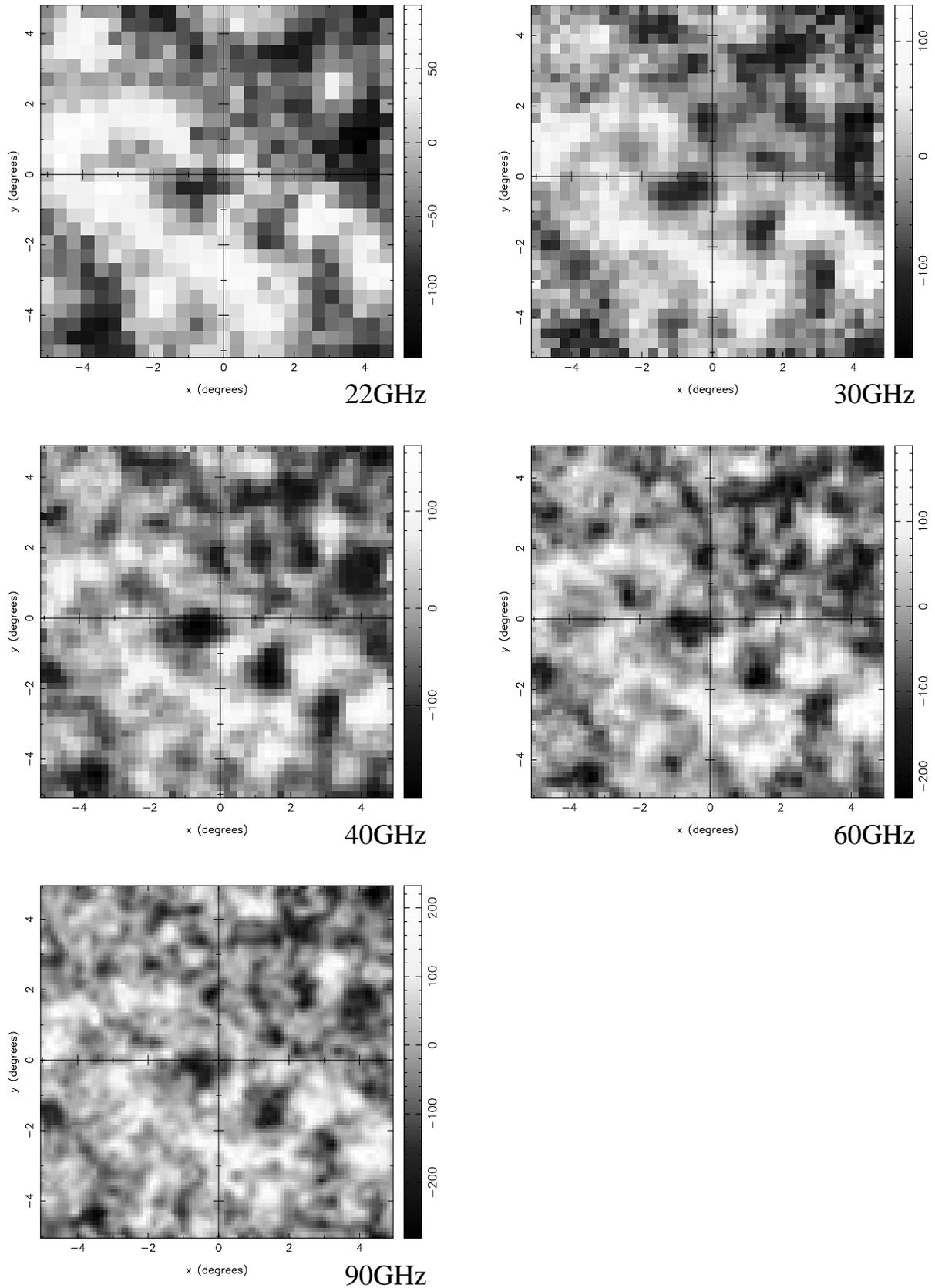,width=16cm}}
\caption{The $10\dg \times 10\dg$ maps observed at each of the five
MAP frequencies listed in Table~\ref{table1}.
At each frequency we assume a Gaussian beam with the appropriate FWHM and
a sampling rate of FWHM/2.4. Isotropic noise with the relevant rms 
has been added to each map. The map units are 
equivalent thermodynamic temperature in $\mu$K.}
\label{fig3}
\end{figure*}

The observed maps at each of the five MAP frequencies are shown in
Figure~\ref{fig3} in units of equivalent thermodynamic temperature
measured in $\mu$K. The coarse pixelisation at the lower observing
frequencies is due to the FWHM/2.4 sampling rate. Moreover, at these
lower frequencies the effect of convolution with the relatively large
beam is also easily seen. It is also seen that, as expected,
the CMB emission dominates each of the frequency channels. 

\section{The component separation}
\label{results}

The component separation is performed
using the Fourier space MEM algorithm developed by HJLB98,
and the reader is referred to that paper for details of the method.
In the absence of non-Gaussian signals the method reduces to
linear Wiener filtering. In this Section, 
we apply the MEM algorithm to the simulated MAP data
discussed in Section~\ref{simmobs}. 
Since the dominant contributions
to the MAP data are the CMB and the pixel noise, which are both
Gaussian, the results of the MEM algorithm should be
very similar to those obtained using a Wiener filter approach 
(Bouchet, Gispert \& Puget 1996; Tegmark \& Efstathiou 1996).
Our primary aim is simply to determine the accuracy
to which the CMB emission can be recovered from MAP observations 
after such a component separation has been performed.

As discussed in HJLB98 there are
several layers of information that we can use in the analysis of the
data. For the CMB and SZ effects the frequency spectra is accurately
known. For the Galactic emissions this is not true. However, as
discussed in Jones et al. (1998b) it is possible to use the data
itself to constrain the spectral dependence of any Galactic emission
that is significant in the data. If it is not a significant component
then the uncertainty in the spectral index will only affect the
reconstruction of that Galactic component and not the reconstruction
of any of the other components. Therefore, in this letter, we assume
perfect knowledge of the frequency spectra of each component.

Our other prior knowledge concerns the spatial distribution, or power
spectrum of the various emission components. We are never entirely
ignorant of the shape of the power spectra for the foregrounds 
in the data. In HJLB98 two levels of power spectrum
information were used. The first assumed no information on the power
spectra of any component and the second assumed perfect knowledge of
the power spectra and cross-correlation information. In this letter we
will assume a more conservative approach than the latter but also
remembering that we have some prior information on the components. We
choose to use the best guess theoretical power spectrum for each
component. This is a white noise power spectra for the two SZ effects,
standard CDM for the CMB and an $\ell^{-3}$ power spectra for the
Galactic components. In any case, as noted in HJLB98, the MEM
algorithm is an
iterative process and the initial guess for the power spectra does
not greatly affect the final reconstructions. 

\subsection{The reconstructed CMB map}
\label{recicf}

\begin{figure}
\centerline{\epsfig{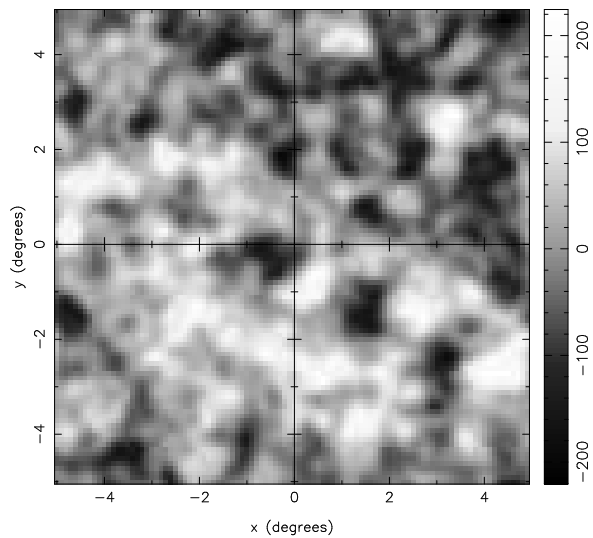}}
\caption{MEM reconstruction of the $10\dg \times 10\dg$ CMB map 
shown in Fig.~\ref{fig1}, using the theoretical power spectra
information (see text). 
The map is plotted at 50 GHz and has been convolved
with a Gaussian beam of FWHM equal to 12.6 arcmin. The map units are
equivalent thermodynamic temperature in $\mu{\rm K}$.}
\label{fig4}
\end{figure}

The reconstruction of the CMB component is shown in
Figure~\ref{fig4}. The greyscale in this figure was chosen to coincide
with that of Fig.~\ref{fig1} in order to enable a more
straightforward comparison with the true map. At least by eye the CMB
has been reproduced quite accurately. The reconstructions of the
other physical components are significantly worse and are not plotted.
In fact, the only other
components for which the reconstruction differs significantly from
zero are the free-free and synchrotron Galactic components. This is
expected as the other three components are well below the noise (see
Figure~\ref{fig2}) in all frequency channels. The free-free and
synchrotron reconstructions appear smoothed to a larger degree than
the CMB reconstruction which is also expected because the two Galactic
components only appear above the noise in the lowest resolution
(22~GHz) channel. Further tests on the thermal SZ effect show that
only clusters with an integrated decrement of $> 250\mu$K (at 22~GHz) will be
observable with the MAP satellite and even then the reconstructed
decrement is severely underestimated. This is because the
frequency dependence of the thermal SZ at frequencies below 100~GHz
closely follows that of the CMB and it is very difficult to extract
the information on the SZ effect from the data.

In order to obtain a quantitative description of the accuracy of the
reconstructions, we calculate the
rms of the residuals. For any
particular physical component, this is given by 
\begin{equation}
e_{\rm rms} = \left[ {1\over N} \sum_{i=1}^N \left( T^{\rm rec}_i -
T^{\rm true}_i \right)^2 \right]^{1/2}
\label{eq:erms}
\end{equation}
where $T^{\rm rec}_i$ and $T^{\rm true}_i$ are respectively the reconstructed
and true temperatures in the $i$th pixel and $N$ is the total number of
pixels in the map. The value of $e_{\rm rms}$ for the CMB
reconstruction is $22\mu$K per $12.6\arcmin$ pixel. 
We note that this is very close to the
desired $20\mu$K accuracy generally quoted for the MAP satellite.
This error should be
contrasted with the result obtained by using the 90 GHz data channel
as the CMB map. In this case,
the error on the map is $46\mu$K (of which $45\mu$K is due
to instrumental noise and $10\mu$K is due to the presence of the unsubtracted 
foregrounds), so some form of component separation is clearly desirable.

As mentioned above, the other physical components were 
only poorly reconstructed. Indeed no recovery of the dust or either SZ
effect was possible at all, and only a
low-resolution reconstructions of the other two components were
obtained. The rms errors on the free-free and synchrotron
reconstructions were $0.6\mu$K and $0.07\mu$K respectively. By
comparing these rms errors
with the peak amplitudes in each map, which are $1.0\mu$K and
$0.10\mu$K respectively, we see that the reconstructions
are not very accurate.

\begin{figure}
\centerline{\epsfig{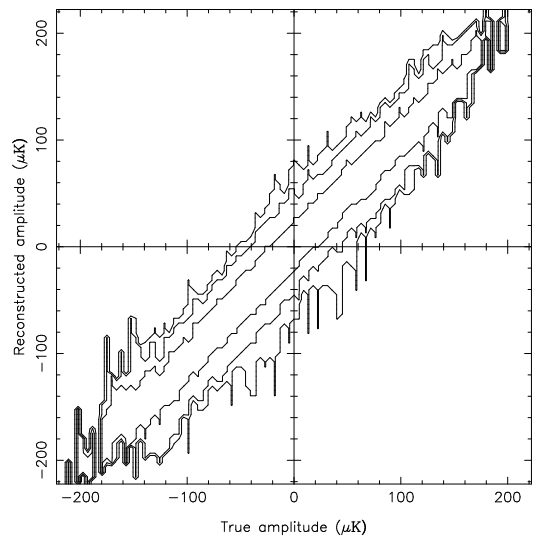}}
\caption{Comparison of the CMB true map with the reconstructed map
using the MEM algorithm.
The horizontal axes show the true map amplitude within a pixel 
and the vertical axes show the reconstructed amplitude. 
The contours contain 68, 95 and 99 per cent of the pixels respectively.}
\label{fig5}
\end{figure}

\begin{figure}
\centerline{\epsfig{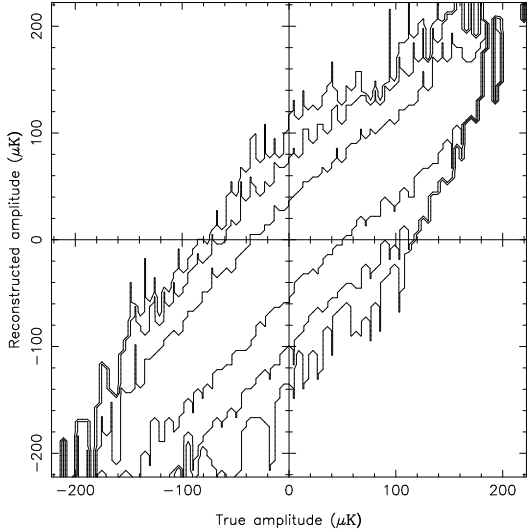}}
\caption{As for Figure~\ref{fig5} but using the 90~GHz data channel as
the CMB `reconstruction'. }
\label{fig6}
\end{figure}

As a further test of the quality of the CMB reconstruction,
we plot the amplitudes of
the reconstructed pixel temperatures against those of the true map. 
The temperature
range of the true input map is divided into 100 bins. Three contours
are then plotted which correspond to the 68, 95 and 99 per cent points
of the distribution of corresponding reconstructed temperatures in
each bin. Clearly, a perfect reconstruction would be represented by a
single diagonal contour of unit gradient with width equal to the bin
size. Figure~\ref{fig5} shows this
comparison for the CMB reconstruction from the MAP data. The gradient
of the best fit line is 1.03 and the 68 per cent contours lie approximately
$25\mu$K on either side of the true value. This agrees with the value
for the $e_{\rm rms}$ quoted above. Figure~\ref{fig6} shows the same
comparison for the case in which the highest frequency data channel as 
used as the CMB reconstruction at $12.6\arcmin$ resolution. We see
that in this case the 68 per cent contour is much wider, corresponding
to about 45 $\mu$K errors, in agreement with the $e_{\rm rms}$ value
quoted above. 

\subsection{The reconstructed CMB power spectrum}

\begin{figure}
\centerline{\epsfig{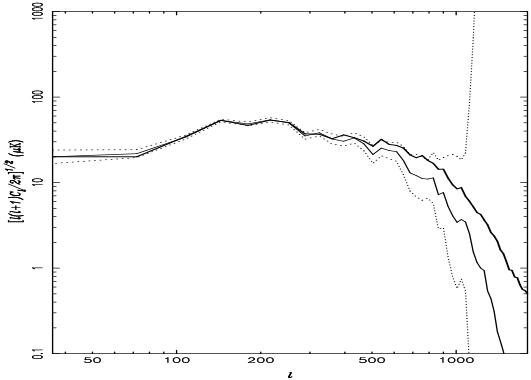}}
\caption{Reconstructed power spectrum, in a $10\dg \times 10\dg$ field, 
for the CMB simulation (bold
line) compared to the true power spectrum (faint line). The 68 per
cent confidence limits (dotted lines) are also shown.}
\label{fig7}
\end{figure}

Since the component separation is performed in the Fourier domain, it
is straightforward to compute the reconstructed power spectrum of the
CMB component. As noted in HJLB98 it is also straightforward to
calculate the errors on the reconstructed power spectrum using the
inverse Hessian. Figure~\ref{fig7} shows the power spectrum (bold
line) of the reconstructed CMB map compared to the power spectrum
(faint line) of the input map. The 68 per cent confidence limits on
the power spectrum reconstruction are also shown (dotted lines). It is
seen that the true power spectrum is almost always contained within
the 68 per cent confidence intervals. These confidence limits on the
power spectrum are calculated by assuming a Gaussian profile for the
posterior probability distribution around its maximum (see HJLB98).
We see that the power spectrum of the reconstructed map is reasonably
accurate for $\ell \la 500$, at which point the reconstructed
spectrum begins slightly to underestimate the true value, and beyond
$\ell \approx 1000$ the input power spectrum is severely
underestimated. This behaviour is due mainly to the effects of beam
convolution and presence of instrumental noise. By using some form of
rescaled filter, it is possible to boost the reconstructed power
spectrum at high $\ell$ so that it lies closer to the input spectrum,
but only at the cost of increasing the rms error of the corresponding
reconstructed CMB map (see HJLB98). Therefore, such a procedure has
not been carried out here.

\section{Discussion and conclusions}
\label{conc}

In this paper, we have analysed simulated MAP observations in a $10\dg
\times 10\dg$ field, in order to separate the CMB emission from
various foreground contaminants and determine how accurately the CMB
emission can be recovered. The algorithm used to perform the component
separation is the Fourier space maximum-entropy method discussed by
Hobson et al. (1998), which reduces to a Wiener filter in the absence
of non-Gaussian signals.  The simulated observations include
contributions from primary CMB, kinetic and thermal SZ effects, and
Galactic dust, free-free and synchrotron emission.  Assuming knowledge
of both the one-dimensional ensemble average power spectrum for each
component and its frequency behaviour, we find that the CMB emission
can be reconstructed with an rms accuracy of $22\mu$K per 12.6 arcmin
pixel.  We note that this represents a substantial improvement as
compared to the individual temperature sensitivities of the raw data
channels, and thus indicates that some form of component separation is
desirable for MAP observations.  We also find that the power spectrum
of the reconstructed CMB map lies close to the true power spectrum for
$\ell \la 1000$. In contrast to our analysis of simulated Planck
Surveyor observations (Hobson et al. 1998), we find that it is not
possible to recover the emission from Galactic dust or the
kinetic/thermal Sunyaev-Zel'dovich effects, although low-resolution
reconstructions of the Galactic free-free and synchrotron emission
were obtained.

We have also repeated the component separation using a straightforward
Wiener filter algorithm, and find that, as expected, the
reconstructions are very similar to those presented above and the rms
error on the reconstructed CMB map is almost identical.  Moreover,
both techniques take the same computational time as they are based
upon similar calculations in Fourier space.  As mentioned above, the
similarity of the results is due to the fact that the dominant
contributions to the MAP data are the CMB and instrumental noise,
which are both Gaussian.  Thus, although several non-Gaussian
foregrounds were included in the simulated observations, the
angular resolution and frequency coverage of the MAP satellite
preclude the presence of significant non-Gaussian effects in the data
(although this is certainly not the case for the Planck Surveyor satellite).
If, however, the input CMB component in the MAP simulations is
replaced by a realisation of CMB fluctuations due to cosmic strings,
there still exists a significant non-Gaussian signal in the data. In
this case we find that the rms error on the CMB reconstruction is
consistently $2\mu$K lower for the MEM algorithm as compared to the
Wiener filter.

\section*{Acknowledgements}

AWJ acknowledges King's College, Cambridge,
for support in the form of a Research Fellowship. We would like to
thank Francois Bouchet and Richard Gispert for kindly providing the
simulations used in this paper.

\bsp  
\label{lastpage}
\end{document}